\documentclass[reprint,prb,aps,citeautoscript,floatfix]{revtex4-1}
\pdfoutput=1
\usepackage{color}
\usepackage{amsmath}
\usepackage{mathrsfs}
\usepackage{xr-hyper}
\usepackage{graphicx}
\usepackage{siunitx}

\newcommand{\UO}{Department of Physics, University of Oregon, Eugene, OR 97403, USA}
\newcommand{\UCSD}{Center for Memory and Recording Research,
University of California, San Diego, CA 92093, USA}
\newcommand{\DECE}{Department of Electrical and Computer Engineering, University of California, San Diego, La Jolla, CA 92093, USA}
\newcommand{\NCEM}{National Center for Electron Microscopy, Molecular Foundry, Lawrence Berkeley National Laboratory, 1 Cyclotron Road, Berkeley, California 94720, USA}

\renewcommand{\d}[1]{\ensuremath{\operatorname{d}\!{#1}}}
\newcommand{\df}{\Delta\!f}
\newcommand{\qp}{\mathbf{q}_\perp}
\newcommand{\qpm}{\lvert\mathbf{q}_\perp\rvert}
\newcommand{\rp}{\mathbf{r}_\perp}
\newcommand{\nablap}{\nabla_\perp}

\begin{document}

\title{A streamlined approach to mapping the magnetic induction of skyrmionic materials}
\author{Jordan J. Chess$^1$}
\author{Sergio A. Montoya$^{2,3}$}
\author{Tyler R. Harvey$^1$}
\author{Colin Ophus$^4$}
\author{Simon Couture$^{2,3}$}
\author{Vitaliy Lomakin$^{2,3}$}
\author{Eric E. Fullerton$^{2,3}$}
\author{Benjamin J. McMorran$^1$}
\email{mcmorran@uoregon.edu}
\affiliation{$^1$ \UO}
\affiliation{$^2$ \UCSD}
\affiliation{$^3$ \DECE}
\affiliation{$^4$ \NCEM}

\date{\today}

\begin{abstract}
Recently, Lorentz transmission electron microscopy (LTEM) has helped researchers advance the emerging field of magnetic skyrmions.
These magnetic quasi-particles, composed of topologically non-trivial magnetization textures, have a large potential for application as information carriers in low-power memory and logic devices.
LTEM is one of a very few techniques for direct real space imaging of magnetic features at the nanoscale.
For Fresnel-contrast LTEM, the transport of intensity equation (TIE) is the tool of choice for quantitative reconstruction of the local magnetic induction through the sample thickness. Typically this analysis requires collection of at least three images.
Here we show that for uniform thin magnetic films which includes many skyrmionic samples, the magnetic induction can be quantitatively determined from a single defocused image using a simplified TIE approach.

\end{abstract}

\maketitle

\section{Introduction}

Magnetic skyrmions are particle-like solitons or magnetic bubbles in a magnetization texture that have topologically non-trivial spin textures\cite{nagaosa_topological_2013}.
The stability of skyrmions and the low current density necessary to move them\cite{yu_room-temperature_2016} has inspired many suggested applications that employ skyrmions as bits in both memory and logic devices which are predicted to be highly energy-efficient\cite{tomasello_strategy_2014, zhang_magnetic_2015, zhang_skyrmion-skyrmion_2015, krause_spintronics:_2016, zhang_thermally_2016, woo_observation_2016}.
These magnetic quasi-particles were initially identified only at low temperatures in non-centrosymmetric crystals including MnSi\cite{muhlbauer_skyrmion_2009, yu_real-space_2010}, FeCoSi\cite{munzer_skyrmion_2010} and FeGe\cite{yu_near_2011}, but recent observations have shown that skyrmions can be stabilized in a more diverse class of materials including those with perpendicular magnetic anisotropy (PMA)\cite{yu_magnetic_2012, yu_biskyrmion_2014, morikawa_lorentz_2015, lee_synthesizing_2016, woo_observation_2016, wang_centrosymmetric_2016}.
This larger swath of materials suggests the need for more rapid characterization techniques to both facilitate the efficient search for materials suitable for applications in skyrmionic devices and explore the basic physics of these magnetic textures.

Lorentz transmission electron microscopy (LTEM) is one of a very few techniques for providing direct real space images of magnetic features at the nanoscale.
Recent improvements in aberration correction and instrument stability have led to a new resolution benchmark of 1 nm for scanning LTEM\cite{mcvitie_aberration_2015}.
Additionally, new tomographic reconstruction algorithms have led to the demonstration of 3D vector field electron tomography by Phatak \textit{et al.}~\cite{phatak_three-dimensional_2010}

Most of the LTEM studies of skyrmion materials have employed analysis based on the transport of intensity equation (TIE),\cite{teague_deterministic_1983, marc_de_graef_magnetic_2001} an equation that relates the \textit{z-}derivative of the image intensity to the phase shift of an electron.
This approach yields quantitative maps of the local in-plane magnetic induction integrated through the sample thickness, but requires multiple images (under-, in-, and over-focused) be taken at a specific point of interest in the sample\cite{marc_de_graef_magnetic_2001}.
In a post-processing step these images are first aligned and then used to approximate the \textit{z-}derivative of the image intensity.
In order to maximize the final field of view, the microscopist must carefully align the microscope to minimize image movement between images recorded as different focus values.
These alignments can be sensitive to changes in other experimental parameters including magnetic field applied to the sample.
This, coupled with the need to properly align images which can be difficult to automate\cite{koch_towards_2014}, increases the total time needed to extract useful information from a magnetic sample.
This often makes certain experiments prohibitively time-consuming, such as determining the in-plane magnetic induction during an \textit{in-situ} applied field sweep (although this type of study does exists in the LTEM literature\cite{budruk_situ_2011, budruk_situ_2011-1}).
An alternative approach is to forego mapping the magnetic induction and instead answer questions that depend only on the location of domain walls, which can in general be accomplished with a single defocused image.
This method has been used to determine the non-adiabatic spin torque parameter\cite{pollard_direct_2012}, image domain wall nucleation\cite{park_4d_2010}, and record skyrmion motion\cite{yu_skyrmion_2012}.
Additionally, Phatak \textit{et al.}, showed that both the polarity and chirality of a vortex magnetization pattern of a magnetic disk can be determined from a single Fresnel contrast image of a tilted sample\cite{phatak_determination_2009}.

\begin{figure}
    \includegraphics[width=8.5cm]{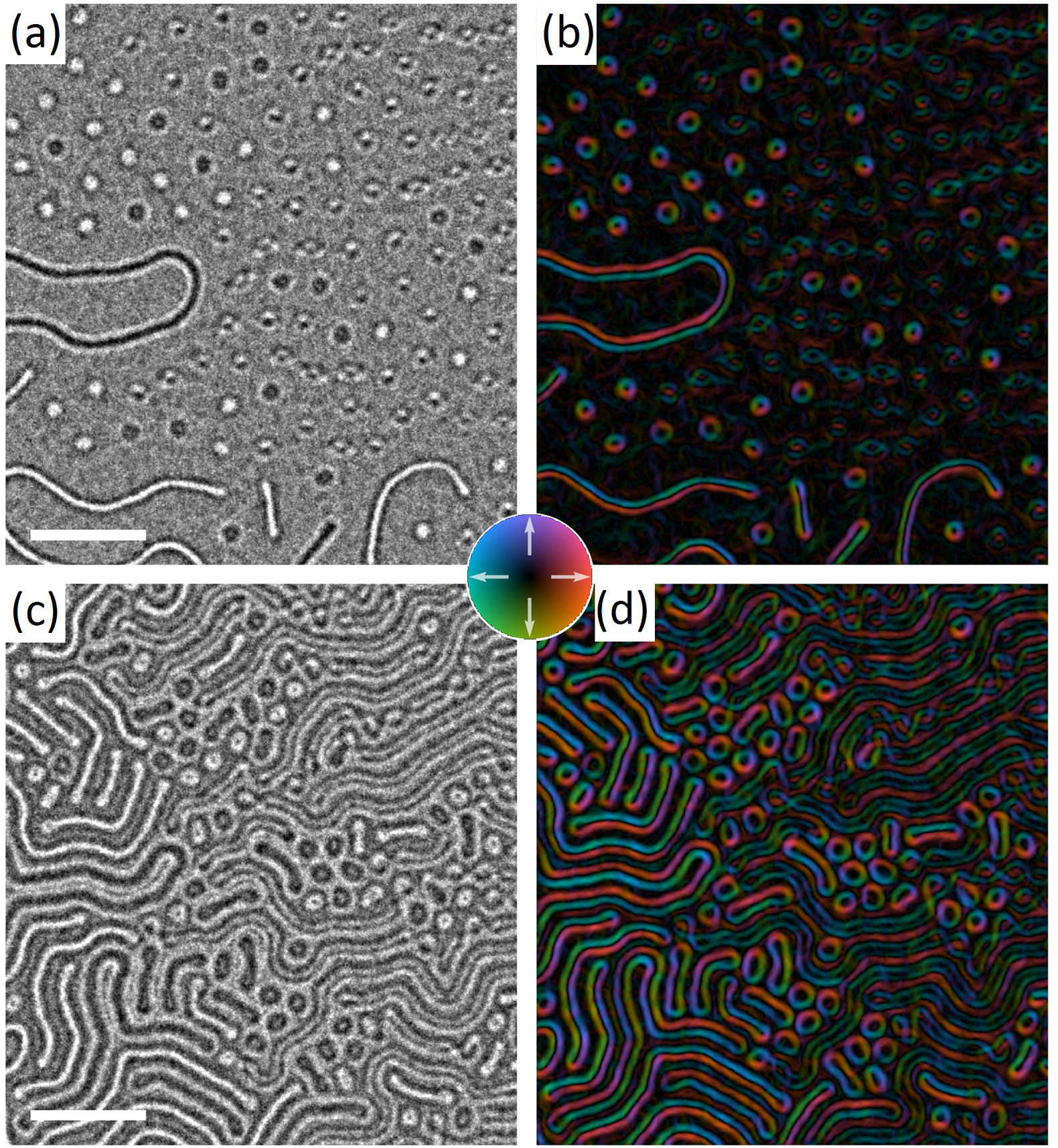}
        \caption{(a) and (c) Selected under-focused Lorentz TEM images from a field sweep performed on a FeGd multilayered thin film with (a, b) \SI{180}{\milli\tesla} and (c, d) \SI{70}{\milli\tesla} field applied perpendicular to the film.
        Scale bar is \SI{1}{\micro\meter}.
        (b), (d), The magnetic induction calculated using our single image analysis on the image to the left (hue and saturation of color indicate the direction and magnitude of the magnetic induction).
        See supplemental material video for full field sweep.}
\label{fig:experimental_SITIE_fild}
\end{figure}

Similar to the work by Paganin \textit{et al.}\cite{paganin_simultaneous_2002}, in which they showed a thickness map of a homogeneous non-magnetic material could be determined from a single defocused image,
here we show that one defocused image is sufficient to determine the magnetic portion of the electron phase shift of a uniform film.
This allows one to map the magnetic induction without the trade-off of a slower, more involved focal series experiment, making it ideal for \textit{in situ} experiments on suitable samples.
Figure~\ref{fig:experimental_SITIE_fild} shows an application of the single image TIE approach we are discussing here, applied to an FeGd multilayered film\cite{lee_synthesizing_2016, montoya_dipolar-stabilized_2016}, under quasi-dynamic conditions.
The data was taken as an applied perpendicular magnetic field was swept from a field strong enough to saturate the sample to a slightly negative applied field.
The data shows skyrmions (black/white circles), Bloch lines, and bubbles with zero topological charge (elliptically shaped) nucleating as the field reduced in strength and then evolving into skyrmions, and labyrinth domains.
The top two images (a,b) are the under-focus LTEM image and reconstructed magnetic induction with $\df$=-\SI{300}{\micro\meter}, and applied field $H_z$= \SI{180}{\milli\tesla}, while (c) and (d) are the under-focus and magnetic induction at $H_z$= \SI{70}{\milli\tesla}.

Additional algorithms for single-image phase retrieval or exit-wave reconstruction exist but require specific sample geometries such as an isolated object\cite{morgan_direct_2011}, or specific illumination conditions and a diffraction image\cite{martin_direct_2008, martin_practical_2011} which make them not suitable for this type of sample or difficult to implement in a TEM.\@
It is worth emphasizing that the general paradigm for these single image phase retrieval algorithms is to use \textit{a priori} knowledge to simplify the analysis, which in practice usually means restricting oneself to a subset of samples.
In this case we are choosing to restrict our analysis to uniform thin magnetic films.
Utilizing our new approach, the full in-plane magnetic induction can be determined for each image in a quasi-dynamic measurement with no extra experimental requirements and fewer post-processing steps.
This fuller understanding is often required to interpret the LTEM images of the complex magnetization textures present in skyrmionic materials.

 \section{Theory}

\begin{figure*}
    \centering
    \includegraphics[width=\textwidth]{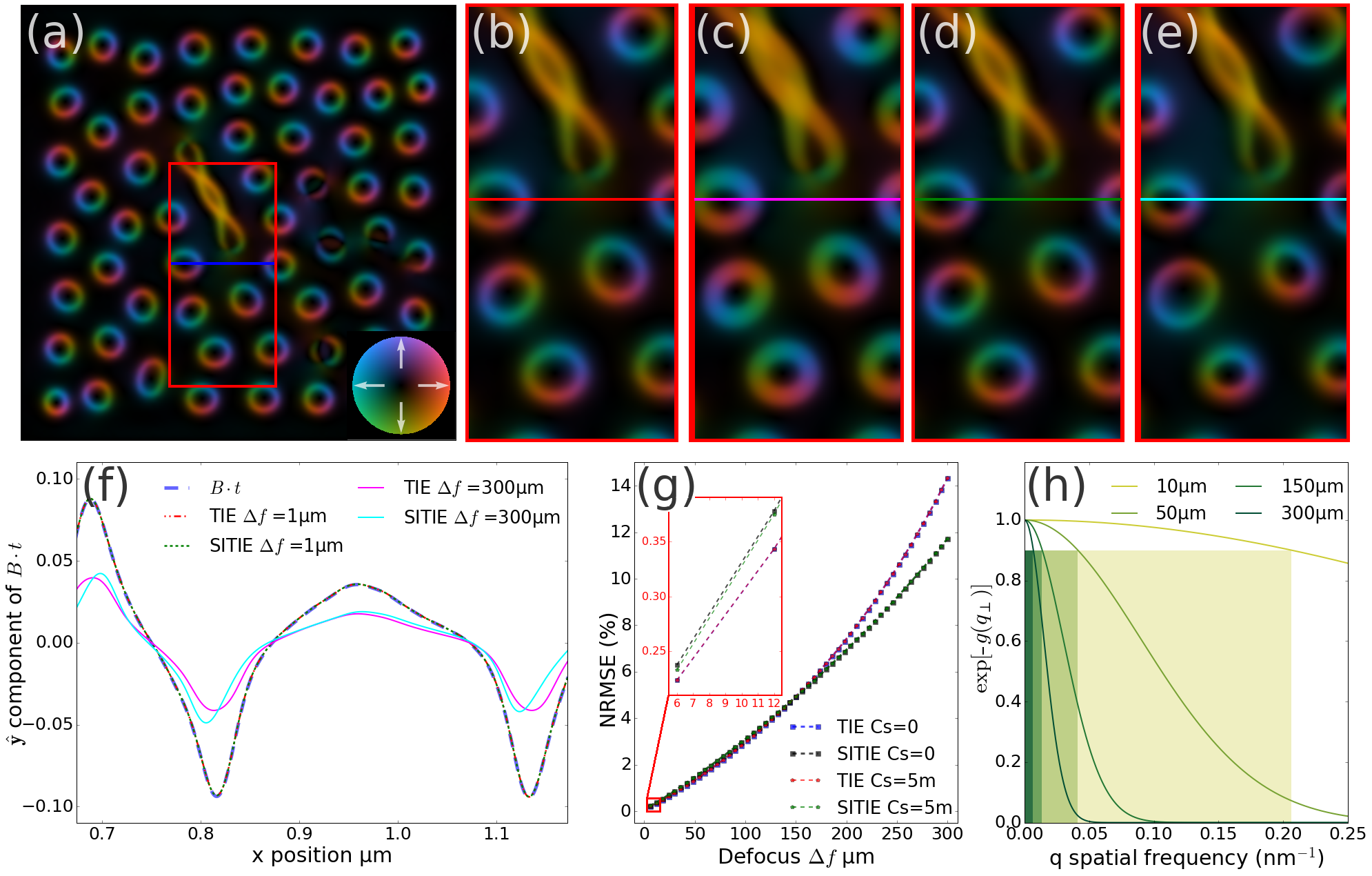}
        \caption{\textbf{Simulations} (a) Plot of in-plane components of the local magnetic induction calculated from output of a micromagnetic simulation (color indicates direction and magnitude of the field).
        (b-e) Local magnetic induction of the region shown in the red box in (a), calculated using conventional TIE with $\df =$ \SI{1}{\micro\meter} and $\df =$ \SI{200}{\micro\meter} for (b) and (c) respectively. SITIE-calculated magnetic induction for (d) $\df =$ \SI{1}{\micro\meter} and (e) $\df =$ \SI{300}{\micro\meter}.
        Notice that only slight distortion errors are present in the $\df =$ \SI{300}{\micro\meter} cases, caused by using a focus outside of the validity of the paraxial approximation.
        (f) y-component of the magnetic induction along the colored lines in images (a-e), note the nearly perfect agreement between the reference, TIE, and SITIE for the $\df=$\SI{1}{\micro\meter}.
        (g) Plot of the total normalized root mean square error in the determination of $B\cdot t$ calculated using equation \eqref{eq:error_metric} as function of defocus for TIE and SITIE showing there is no practical difference between the methods for moderate defocus.
        For these simulations $\Theta_c$ was set at \SI{5 e -5}{\radian}.
        The inset shows that for any $\df$ larger than \SI{10}{\micro\meter} the effects of including a non-zero $C_s$ are truly negligible.
        (h) Plot of $\exp{[-g(q_{\perp})].}$ for various values of $\df$ for a constant $\Theta_c=$ \SI{5 e -5}{\radian}.
        The shaded regions indicate where $\exp{[-g(q_{\perp})]}>0.9$, marking the length scales where SITIE can be comfortably applied.}
\label{fig:1vs3tie}
\end{figure*}

The phase imparted on an electron plane wave traveling along the \textit{z-}axis after transmission through a sample with electric potential $V$ and vector potential $\mathbf{A}$ is given by the Aharonov-Bohm phase shift\cite{aharonov_significance_1959}:
\begin{equation}
\begin{aligned}
\phi(\rp)= & \, C_E \int_L V(\rp,z) \d{z} -
\frac{e}{\hbar}\int_L \mathbf{A}(\rp,z)\cdot \d{\mathbf{r}}\\
\equiv & \, \phi_e+\phi_m
\end{aligned}
\end{equation}
where $L$ is a path parallel to the propagation direction of the electron beam, $\rp$ is the location in the sample plane, $C_E$ is the interaction constant\cite{marc_de_graef_magnetic_2001}, $e$ is the electron charge, and $\hbar$ is the reduced Planck's constant.
If we assume a homogeneous foil of uniform thickness $d$ and constant mean inner potential $(V_0)$ the electrostatic term can be easily evaluated and yields,
\begin{equation*}
\phi_e = C_E V_0 d
\end{equation*}
Additionally, the effects of inelastic scattering and high angle scattering of electrons out of the optical system can be described by an exponential drop in the initial amplitude of the electron wave function.
Thus, assuming parallel illumination, the complex amplitude exiting the foil is,
\begin{equation}
\psi_0(\rp) = Ae^{-\alpha d}e^{iC_E V_0 d}e^{i \phi_m(\rp)}.\label{eq:exit_wave}
\end{equation}
The intensity of the wave at the image plane using the microscope transfer function ($ \mathcal{T}(\qp)$) is then given by,
\begin{equation}
I(\rp, \df) = \lvert \mathscr{F}^{-1} \left \lbrace
\mathscr{F}[\psi_0(\rp)]\mathcal{T}(\qp) \right\rbrace \rvert ^2 \label{eq:a}
\end{equation}
where $\qp$ are the in-plane spatial frequencies.
A relevant transfer function that models the effects of spherical aberration ($C_s$) and a damping envelope ($g(q_{\perp})$) due to a spread in illumination angles caused by lens instabilities is:
\begin{equation}
\mathcal{T}(\qp) = a(\qpm) e^{- i \chi(q_{\perp})}e^{-g(q_{\perp})}\label{eq:transfer}
\end{equation}
where $a(q_{\perp})$ is an aperture function, the phase transfer function $\chi(q_{\perp})$ is described by,
\begin{equation}
\chi(q_{\perp}) = \pi \lambda \df q_{\perp} ^2 + \textcolor{blue}{\frac{1}{2} \pi C_s \lambda^3 {q_{\perp}}^4} \label{eq:phase_transfer}
\end{equation}
and $g(q_{\perp})$ given in terms of the divergence angle $\Theta_c$ is\cite{walton_malts:_2013},
\begin{equation}
g(q_{\perp}) = {\left( \frac{\pi \Theta_c}{\lambda}\right)}^2
{\left(\textcolor{blue}{C_s \lambda^3 q_{\perp}^3} + \df \lambda q_{\perp} \right)}^2 \label{eq:damping}
\end{equation}
Above, $\lambda$ is the relativistic electron wavelength, $\df$ is the distance from the in-focus plane, and we have used $q_{\perp} \equiv \qpm $ for notational convenience.
Before continuing we stop to note that conventional TIE analysis presumes both of the blue terms in equations \eqref{eq:phase_transfer} and \eqref{eq:damping} are negligible.
This is generally a reasonable assumption because of the large defocus values used in LTEM, for example see Figure~\ref{fig:1vs3tie}.
As will be discussed, our method neglects one additional term  (the last term in equation \eqref{eq:damping}).
For completeness and accuracy the full transfer function (eq. \eqref{eq:transfer}) was used in all image simulations.

Taylor expanding the transfer function for small $q_{\perp}$ and small $\df$, the ``paraxial approximation'', we arrive at an approximate form of equation \eqref{eq:a}\cite{graef_quantitative_2001}.
\begin{equation}
\begin{aligned}
 I(\rp, \df) & \approx  I_0 -
 \frac{\lambda \df}{2 \pi} \nablap \cdot (I_0 \nablap \phi_m)\\
 & + \frac{{(\pi \Theta_c\df)}^2}{2\ln{2}}[\sqrt{I_0} \nablap^2 \sqrt{I_0} -
I_0{(\nablap \phi_m)}^2]\label{eq:I1}
\end{aligned}
\end{equation}
here $I_0 = \lvert \psi_0(\rp) \rvert^2$.
Examining equation~\eqref{eq:exit_wave} we see that if we are analyzing homogeneous thin film specimens with a uniform thickness, which includes many materials, then $I_0$ becomes a constant, as shown, for example in Figure\ref{fig:experimental_tie_SITIE}.b.
And equation~\eqref{eq:I1} simplifies to,
\begin{equation}
\begin{aligned}
I(\rp, \df)  \approx  I_0  \biggl( 1 &
- \frac{\lambda \df}{2 \pi} \nablap^2 \phi_m \\
& - \frac{{(\pi \Theta_c\df)}^2}{2\ln{2}}{(\nablap \phi_m)}^2 \biggr)\label{eq:I}
\end{aligned}
\end{equation}
As show by De Graef \textit{et al.}\cite{graef_quantitative_2001} the transport of intensity equation can be obtained from~\eqref{eq:I} by simply subtracting the value at $I(\rp,\pm \df)$ yielding,
\begin{equation}
\begin{aligned}
\nablap^2 \phi_m = & -\frac{2 \pi}{I_0 \lambda} \frac{I(\rp, \df)-
I(\rp, -\df)}{2 \df} \\ \approx &  -\frac{2 \pi}{I_0 \lambda} \frac{\partial I}{\partial z}.\label{eq:TIE}
\end{aligned}
\end{equation}
In this way, the Laplacian of the phase can be derived from two different images of the specimen recorded under different focal conditions. Equation~\eqref{eq:TIE} is the standard equation used in analyzing LTEM data.
Note that a crucial step in standard use of TIE analysis is the calculation of the difference between two images (Eq.~\eqref{eq:TIE} RHS).
Thus, the reconstructed magnetic phase is subject to errors introduced when acquiring images under different conditions including: drift, rotations, and changes in magnification.

Here we suggest a further approximation which can be viewed as an assumption of coherent illumination, such that $(\pi \Theta_c \df q_{\perp})^2 \ll 1$, making the last term in equation \eqref{eq:I} negligible.
That is, this assumes that $\Theta_c$ is small compared to the ratio of the feature size to be resolved over the defocus.
Nature ultimately sets a limit on the highest spatial frequencies that can arise from magnetic features: the inverse of the exchange length, which is on the order of \SI{1}{\nano\meter}$^{-1}$\cite{hubert_magnetic_1998}.
Typical values for $\Theta_c$ used in the literature range from $(1-5) \times 10^{-5}$ radians\cite{marc_de_graef_magnetic_2001, walton_malts:_2013} and, as shown in Figure \ref{fig:1vs3tie}.g, these can be used to set an upper bound on the $\df$ values for which this approximation is valid at roughly \SI{1}{\micro\meter}.
This bound is of course relaxed if the domains present in the sample vary over a larger length scale, as is the case for the data presented here.

This approximation results in a Single Image Transport of Intensity Equation (SITIE),
\begin{equation}
\nablap^2 \phi_m \approx -\frac{2 \pi}{\lambda \df} \left(1 - \frac{I(\rp, \df)}{I_0}
\right).\label{eq:SITIE}
\end{equation}
Essentially by using equation \eqref{eq:SITIE}, one assumes $\exp{\left[-g(q_{\perp})\right]} \approx 1$, anywhere $\mathscr{F}[\psi_0(\rp)]$ has large Fourier components.
One then needs a suitable value of $I_0$.
Here we approximate it as the mean on the defocused image, $I_0 \approx <I(\rp,\df)>_{x,y} \equiv \overline{I}_0$.

Multiple techniques have been developed to solve the standard TIE equation including a Fourier-based approach\cite{paganin_noninterferometric_1998}, a multigrid algorithm\cite{gureyev_hard_1999}, a symmetrized version of the Fourier method\cite{volkov_new_2002}, and finite element method\cite{lubk_transport_2013}, all of which can also be applied to the SITIE to determine the phase of the exit wave.
From this phase the local magnetic induction can easily be determine using the relation,
\begin{equation}
 \nablap\phi_m(\rp) = - \frac{e}{\hbar}[B(\rp)\times
 \hat{\mathbf{e}}_\mathbf{z}]d
\end{equation}
where $\hat{\mathbf{e}}_\mathbf{z}$ is a vector parallel to the beam propagation direction.

\section{Methods}

\subsection{Micromagnetic simulations}
To validate SITIE and quantify the errors associated with this method, we simulated through-focal series images of an exactly known, simulated magnetization textures.
These micromagnetic textures were obtained from Landau-Lifshitz-Gilbert simulations calculated using the FastMag solver\cite{chang_fastmag:_2011}.
The micromagnetic simulation is for a \SI{2}{\micro\metre} $\times$ \SI{2}{\micro\metre} $\times$ \SI{80}{\nano\metre} ferromagnetic film, using experimentally measured values for the saturation magnetization ($M_s$ = \SI{4e5}{\ampere / \meter^3}), anisotropy constant (K = \SI{4e4}{\joule/ \meter^3}), Gilbert damping ($\alpha = 0.05$), and exchange stiffness ($A_{ex}$ = \SI{5e5}{\joule / \meter}).
An applied perpendicular magnetic field of $H_z$ = \SI{0.2}{\tesla} was used, and the system is allowed to relax to an equilibrium state in \SI{10}{\nano\second}.
These parameters result in the in-plane magnetic induction pattern shown in Figure~\ref{fig:1vs3tie}.a.

\subsection{Lorentz Image Simulations}
Fresnel-contrast LTEM images were simulated using the Mansuripur algorithm: the magnetic phase shift imparted on the electron wave by the results of the micromagnetic simulation was calculated and then equation~\eqref{eq:transfer} was used to propagate the wave to a given defocus plane\cite{mansuripur_computation_1991}.
The electrostatic phase shift was neglected in the simulations, in line with the theory above, as it only contributes an overall constant phase and doesn't contribute to the image intensity.
Prior to applying the Mansuripur algorithm the output of the FastMag simulations were expanded from 200$\times$200 arrays to 2048$\times$2048, and then padded with zeros to a total array size of 4096$\times$4096 to mitigate the introduction of any artifacts from the Fourier-based approach used in both the Mansuripur algorithm and transfer function formalism.
The parameters used for image simulations were: accelerating voltage \SI{300}{\kilo\volt}, defocus values $\df=\SI{1}{\micro\meter}-\SI{300}{\micro\meter}$, and spherical aberration $C_s = 0-\SI{5}{\meter}$.
These values more than cover the range encountered in both standard and aberration-corrected microscopes during an LTEM experiment.
The normalized root mean square error is used as an metric to compare the reconstructed phase to the known phase calculated as,
\begin{equation}
\mathbf{NRMSE} = \frac{\sum_{i=0}^{i=1}\sqrt{\frac{\sum_{m,n} \left( (\tilde{\mathbf{B}}t)_{n,m}-(\mathbf{B}t)_{n,m}\right)^2}{nm}}}{(\mathbf{B}t)_{max}-(\mathbf{B}t)_{min}}\times 100 \%
\label{eq:error_metric}
\end{equation}
where $\tilde{\mathbf{B}}$ is the TIE/SITIE reconstructed local magnetic induction, $\mathbf{B}$ is the known magnetic induction, $t$ the sample thickness, $(m,n)$ the array indices, and $(i=0,1)$ the components of the vector.

\section{Evaluation of SITIE}
\subsection{Numerical evaluation}
To quantitatively analyze the validity of SITIE compared to TIE, we numerically simulated Fresnel-contrast images from simulated domain structures obtained from the micromagnetic simulation.
This allows us to compare the two phase retrieval methods in the absence of noise or any misalignments in images that could cause errors in standard TIE analysis.
Additionally, it gives us a known reference to quantify results that is not present when analyzing experimental data.
The Fourier transform-based method of solving the transport of intensity equation was utilized to reconstruct the phase of both the experimental and simulated data\cite{gureyev_phase_1996,marc_de_graef_magnetic_2001}.
A comparison of the two methods applied to experimental data is left to the next Subsection~\ref{ssec:expeval}.

Figure~\ref{fig:1vs3tie}.(b-e), show the calculated local magnetic induction from both TIE (b,c) and SITIE (d,e) each under two different focal conditions; the first (b,d) from a small defocus (\SI{1}{\micro\meter}) and the second from a large defocus value (c,e) (\SI{200}{\micro\meter}).
Notice the close agreement between the reference and both TIE and SITIE for small defocus (FIG.\ref{fig:1vs3tie}.a,b,d), which have a normalized root mean square error (NRMSE) of 0.169 \% and 0.170 \% respectively.
Interestingly, for the large defocus (\SI{300}{\micro\meter}) examples (figure~\ref{fig:1vs3tie}.e,f) the error associated with TIE (14.3 \%) is larger than that for SITIE (11.7 \%).
These results can be understood analytically from the right hand side of equations~\eqref{eq:TIE} and~\eqref{eq:SITIE}.
They are the central and forward difference approximations for the z-derivative of the image intensity, and have associated errors of order $\mathcal{O}(\df^2)$ and $\mathcal{O}(\df)$ respectively.
This quadratic versus linear error is evident in Figure~\ref{fig:1vs3tie}.g.
Also, evident in Figure~\ref{fig:1vs3tie}.g is the well-known fact that for all but the smallest defocus values used in LTEM, the effects of spherical aberration are negligible\cite{marc_de_graef_magnetic_2001}.
\subsection{Experimental evaluation}
\begin{figure}
    \includegraphics[width=8.0cm]{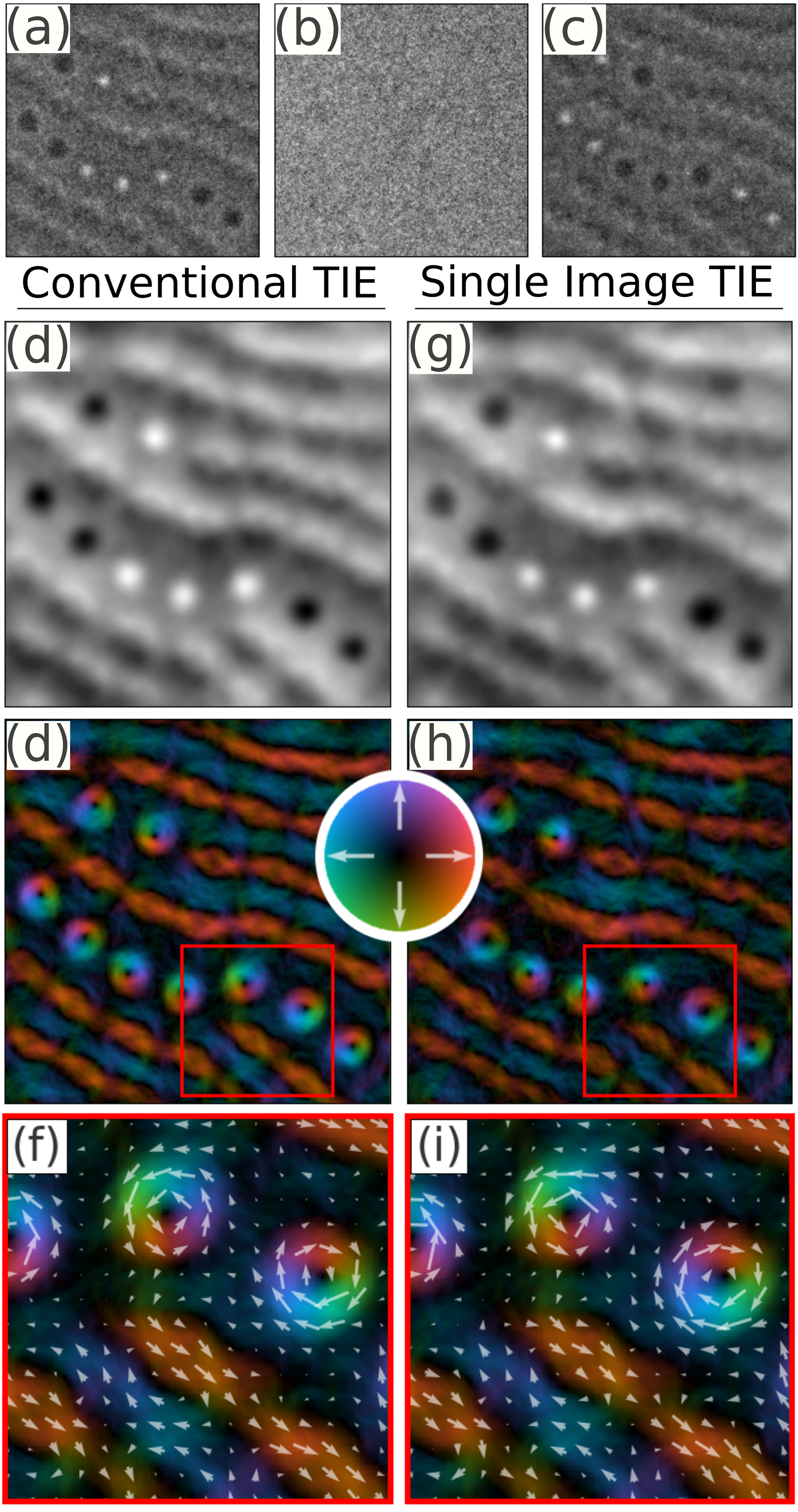}
        \caption{(a-h) Experimental Lorentz TEM analysis of a FeGd multilayered thin film over the same \SI{1.5}{\micro\meter} field of view.
        (a)Under-focused, (b) in-focus, and (c) over-focused images showing Fresnel-contrast((a,c) recorded at $\df=\pm$\SI{300}{\micro\meter}).
        (d) Phase calculated using the standard TIE applied to image (a-c).
        (g) Phase calculated using only image (a). (e), (h), The magnetic induction calculated from phase above.
        (f), (i) Enlarged area from boxed region in (e), (h) with magnetization represented both by color and vector arrows.}
\label{fig:experimental_tie_SITIE}
\end{figure}
\label{ssec:expeval}
To ensure the validity of SITIE on real data, we collected and analyzed through-focal series images of magnetic bubble domains in a thin film sample.
The images were collected using an FEI Titan equipped with a Lorentz lens and integrated CEOS objective lens aberration corrector.
The standard objective lens was partially excited to apply a magnetic field perpendicular to the sample plane.
The sample is nominally a [$\mathrm{Gd}$ (\SI{0.4}{\nano\meter})/$\mathrm{Fe}$ (\SI{0.34}{\nano\meter})] $\times$ 80 multilayered film deposited by DC magnetron sputtering onto \SI{50}{\nano\meter} $\mathrm{Si_3N_4}$ membrane with \SI{20}{\nano\meter} Ta seed and capping layers\cite{lee_synthesizing_2016}.

Prior to analysis all experimental images were filtered following the method suggested by Tasdizen \textit{et al.} to remove low-frequency artifacts caused by slightly non-uniform illumination\cite{tasdizen_non-uniform_2008}.
Figure~\ref{fig:experimental_tie_SITIE} shows the focal series (a-c) for $\df=$(-300,0,\SI{300}{\micro\meter}).
The left column shows (d) the phase reconstructed using conventional TIE analysis applied to (a-c), (e) the magnetic induction determined using the phase in (d) represented with color indicating the magnitude and direction of the magnetic induction and (f) giving a closer look at the region inside the red square in (e).
The right column shows the phase (g) and magnetic induction (h,i) all determined using only image (a).
Included in the images are skyrmions, four of which have helicity $ \gamma = \pi / 2 $ (white circles in phase images), and five with $ \gamma = -\pi / 2 $ with $\gamma$ defined the same as equation B3 in reference (\cite{nagaosa_topological_2013}).
The remaining features are stripe domains starting to break up into topologically trivial bubbles, and skyrmion bound pairs\cite{lee_synthesizing_2016}.
It is important to note that the slightly lower signal-to-noise present in Figure~\ref{fig:experimental_tie_SITIE}.g is not an inherent difference between SITIE and TIE, but instead a consequence of Figure~\ref{fig:experimental_tie_SITIE}.g having half the effective exposure time due to it being calculated from only one image.
This could easily be overcome by increasing the exposure time for images collected for SITIE, or by collecting multiple shorter exposures images aligning and averaging them latter.
We emphasize here that aligning images collected at the same focus value can be accomplished using simple algorithms such as cross-correlation and is significantly easier than aligning images at different foci, because of the associated reversals in contrast, rotation, and distortions between images.
Errors in image alignment caused by pixel shift, magnification changes, and rotations can cause significant errors in the reconstructed phase when performing TIE analysis.
For a detailed discussion of this subject we refer the reader to chapter 5.3.2 of De Graef and Zhu~\cite{marc_de_graef_magnetic_2001}.
SITIE is free of all these errors.

\section{Conclusion}
We have demonstrated, both numerically and experimentally, that a single Lorentz TEM image can appropriately be used to map the magnetic phase of uniform samples, specifically for thin films exhibiting skyrmionic phase.
This simplified TIE approach gives roughly equivalent results to conventional TIE analysis.
Using SITIE analysis on uniform samples simplifies both the computational load and data collection involved in characterizing topological magnetization textures.
Furthermore, this simplification opens the door to exploring new phenomenon that was previously impractical with the traditional TIE analysis by: removing the need to align and collect multiple images, and reducing errors caused by distortions in images.
This simplified technique allows for phase reconstruction during quasi-dynamic measurements (e.g.\ field and/or temperature sweeps), and gives a potential route to ultra-fast LTEM studies.

\section{Acknowledgments}
This work was partially supported by the U.S. Department of Energy, Office of Science, Basic Energy Sciences, under Award No. DE-SC0010466 (JJC, TRH, BJM).
CO acknowledges that work at the Molecular Foundry was supported by the Office of Science, Office of Basic Energy Sciences, of the U.S. Department of Energy under Contract No. DE-AC02- 05CH11231.
Work at UCSD including materials synthesis and characterization was supported by U.S. Department of Energy (DOE), Office of Basic Energy Sciences (Award No. DE-SC0003678).
The authors wish to thank Josh Razink of the CAMCOR facility at University of Oregon. The CAMCOR High-Resolution and Nanofabrication Facility (TEM, FIB and SEM) are supported by grants from the W.M Keck Foundation, the M.J. Murdock Charitable Trust, ONAMI, the Air Force Research Laboratory (agreement number FA8650--05--1--5041), NSF (award numbers 0923577, 0421086) and the University of Oregon.

\section{References}
\bibliographystyle{apsrev4-1}
\bibliography{bibtex}{}

\end{document}